\documentclass[12pt]{iopart}
\usepackage[latin1]{inputenc}
\usepackage[english]{babel}
\usepackage{graphicx}

\begin{document}

\pagestyle{plain}

\title{Microscopic versus macroscopic approaches to non-equilibrium systems}

\author{Bernard Derrida}
\address{Laboratoire de Physique Statistique, 
Ecole
Normale Sup\'erieure,
\\
UPMC Paris 6, Universit\'e Paris Diderot Paris 7, CNRS,
\\24, rue Lhomond, 75231 Paris Cedex 05,
France}

\date{\today}

\begin{abstract}
The one dimensional symmetric simple exclusion process (SSEP) is one of the very few exactly soluble models of non-equilibrium statistical physics. 
It describes a system of particles  which  diffuse with hard core repulsion on a one dimensional lattice in contact with two reservoirs of particles at unequal densities.
The goal of this note is to   review the two main approaches which lead to the  exact expression of  the large deviation  functional of the density of the SSEP in its steady state:
a microscopic approach (based on the matrix product ansatz and
an additivity property) and a macroscopic approach (based on the macroscopic fluctuation theory of Bertini,   De Sole,  Gabrielli,  Jona-Lasinio and  Landim). 
\end{abstract}

\section{The   Symmetric Simple Exclusion Process }
\label{intro}
Understanding   the steady state  properties of systems in contact with two heat baths at unequal temperatures
or two reservoirs of particles at unequal densities is a central question in the theory of non-equilibrium systems \cite{BLR,LLP,EPR1,ST}.
Here I would like to focus on  one exact result 
  which   was obtained during the  last decade on the steady state 
 of one of the simplest models of a non-equilibrium system, the one dimensional symmetric simple exclusion process (SSEP).
For the SSEP this exact result, which gives an   expression of the large deviation functional of the density,
  can be derived either from a microscopic description of the steady state
\cite{DLS1,DLS2,derrida-phys-rep}
 or from a macroscopic approach, which was developed 
by  Bertini,   De Sole,  Gabrielli,  Jona-Lasinio and  Landim 
\cite{BDGJL1,BDGJL2,BDGJL2a}.
These two approaches are discussed below.

\begin{figure}[ht]
\centerline{\includegraphics[width=8cm]{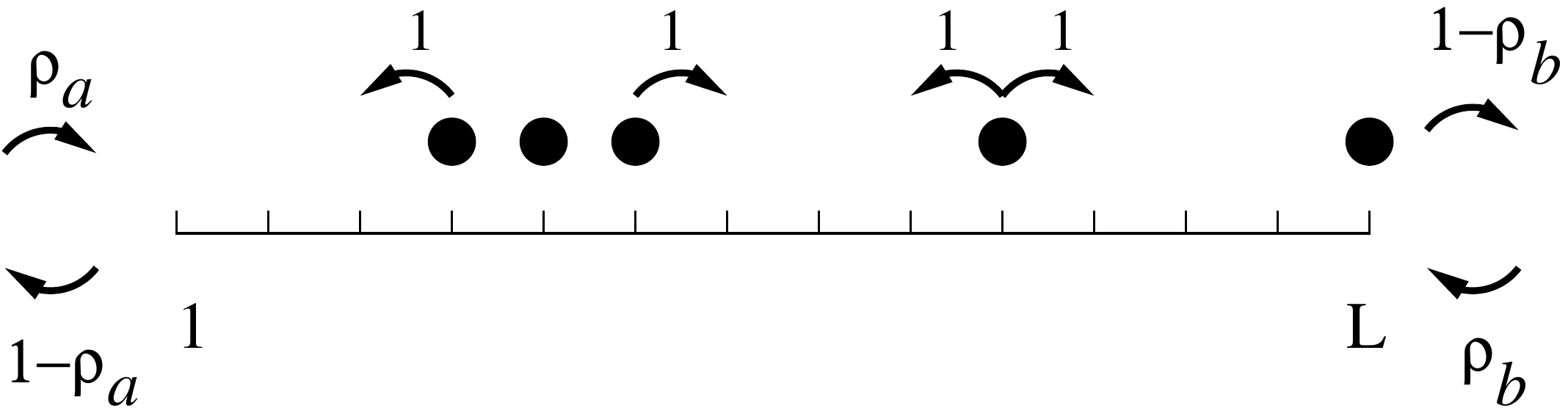}}
\label{ssep}
\caption{The symmetric simple exclusion process: particles diffuse with hard core repulsion on a one dimensional lattice 
connected at its ends to two reservoirs of particles at densities $\rho_a$ and $\rho_b$. }
\end{figure}
The symmetric simple exclusion process (SSEP)  describes a gas of
particles diffusing on a lattice with an exclusion rule which prevents a
particle to move to a site already occupied by another particle 
\cite{Richards,HS,Liggett,KL,SS}. Here we
consider the  one dimensional version with open boundaries. The lattice
consists of $L$ sites,
 each site being either
occupied by a single particle or empty.  During every infinitesimal time
interval $dt$, each particle has a probability $dt$ of jumping to the
left if the neighboring site on its left is empty, $dt$ of jumping to the
right if the neighboring site on its right is empty. At the two
boundaries the dynamics is modified to mimic the coupling with reservoirs
of particles at densities $\rho_a$ for the left reservoir and $\rho_b$ for the right reservoir: at the left boundary, during each time interval $dt$, a
particle is injected on site $1$ with probability $ 2 \rho_a dt$ (if this
site is empty) and a particle is removed from site $1$ with probability
$ 2(1 - \rho_a) dt$ (if this site is occupied). Similarly on site $L$, particles
are injected at rate $2 \rho_b$ and  removed at  rate $2(1-\rho_b)$.
(The factors 2 in the boundary rates simplify  some  expressions below but do not affect the large scale properties).

The SSEP is obviously  a model of transport of particles between two reservoirs at densities $\rho_a$ and $\rho_b$. 
It  is also a simple model of heat transport
between two heat baths at temperatures $T_a$ and $T_b$,  if one interprets each particle as a quantum of energy $\epsilon$, with 
\begin{equation}
\exp \left[ -{\epsilon \over k T_a}\right] = {\rho_a \over  1-\rho_a} \ \ \ \ ; \ \ \ \ 
\exp \left[ -{\epsilon \over k T_b}\right] = { \rho_b \over  1-\rho_b}  \; .
\label{TaTbdef}
\end{equation}

Under the evolution rules of the SSEP, the system reaches, in the long time limit, a steady state. 
If one divides the system of length $L$ into $n$ boxes of size $l$ (with of course $L= n l$),
one can try to determine, in this steady state,  the probability of a certain density profile $\left\{ r_1, r_2 .. r_n \right\}$, i.e. the probability of seeing  $ l r_1 $ particles in the first box, $l r_2$ particles in the second box, ... $l r_n $ particles in the $n$th box.
For large $L$, one expects the following $L$ dependence of this probability
\begin{equation}
{\rm Pro}_{L}(r_1,... r_n|\rho_a,\rho_b) \sim \exp[ - L {\cal F}_{n}(r_1, r_2, ...r_n |\rho_a,\rho_b)]
\label{finite}
\end{equation}
where ${\cal F}_{n}(r_1, r_2, ...r_n|\rho_a,\rho_b)$ is  called the  large deviation function \cite{Touchette} of the density profile $\left\{ r_1, r_2 .. r_n \right\}$.  
When the number $n$ of boxes becomes large, keeping the number $l$ of sites in each box also large, one  can introduce a continuous variable $x=k/n$,   the densities $r_1, r_2 .. r_n $  become  a density profile
$$\rho(x) = \rho\left({k \over n}\right)= r_k$$ and the  large deviation function $ {\cal F}(r_1, r_2, ...r_n |\rho_a,\rho_b)$
 becomes a functional ${\cal F}(\{\rho({x})\}|\rho_a,\rho_b)$  of  the  density profile $\rho({x})$
\begin{equation}
{\rm Pro}_{L}(\{\rho({x})\}|\rho_a,\rho_b) \sim \exp[ - L {\cal F}(\{\rho({x})\} |\rho_a,\rho_b)]  \; .
\label{continu}
\end{equation}
The result which was obtained in 
\cite{DLS1,BDGJL2,DLS2} is  that,  
in the non-equilibrium case, i.e. for $\rho_a \neq\rho_b$,  the exact expression of the large deviation function ${\cal F}(\{\rho({x})\}|\rho_a,\rho_b)$ is given by
\begin{equation}
{\cal F}(\{\rho(x)\}|\rho_a,\rho_b)= \int_0^1 dx  \ \left[ B(\rho(x),F(x)) + \log{F'(x)
\over \rho_b - \rho_a}  \right]
\label{F2}
\end{equation}
where $B(\rho,r)$ is given by 
\begin{equation}
B(\rho,r) = (1- \rho) \log {1- \rho \over 1 - r} \  +\  \rho \log {\rho
\over r}
\; 
\label{Br}
\end{equation}
and where the function $F(x)$ is the monotone solution of the differential
equation
\begin{equation}
\rho(x) = F + {F(1-F) F'' \over F'^2 }
\label{F3}
\end{equation}
satisfying the boundary conditions $F(0)= \rho_a$ and $F(1)= \rho_b$.

\section{Properties of this large deviation function}

Before describing the two main approaches which led to  (\ref{F2},\ref{Br},\ref{F3}), let us discuss briefly a few  properties of the  functional ${\cal F}$.

 One can solve perturbatively (\ref{F3}) 
for $\rho_a$ close to $\rho_b$ and get
\begin{eqnarray}
 \fl F  =    \rho_a - (\rho_a- \rho_b) x
\label{rhoa-rhob-small}
- {(\rho_a - \rho_b)^2 \over \rho_a (1-\rho_a)}
  &  \left[   (1-x) \int_0^x  y \; (\rho(y)- \rho_a) \; dy  \right.
\\ 
 &   \ \ \ \ \left.   + x \int_x^1 (1-y)\;  (\rho(y)-
\rho_a)\;  dy  \right] + O \left((\rho_a-\rho_b)^3\right)
\  .  
\nonumber 
\end{eqnarray}
Therefore in the limit $\rho_b \to \rho_a$, the expression (\ref{F2}) reduces to\begin{equation}
{\cal F}(\{\rho(x)\}|\rho_a,\rho_a)= \int_0^1 dx  \ \left[ B(\rho(x),\rho_a) 
  \right]
\label{F4}
\end{equation} 
This is not surprising as, when $\rho_a=\rho_b$,  the dynamics satisfies  detailed balance  and  in the steady state  all the lattice sites are occupied independently with probability $\rho_a$ (Bernoulli measure).
In this equilibrium case  ${\cal F}$
is  a {\it local}  functional (\ref{F4}) of the density profile $\rho(x)$. This  is a special case of the much more general fact \cite{derrida-phys-rep,BDGJLb} that,  for any system  (with short range interactions) {\it at equilibrium}, in contact with one or several reservoirs  at density $\rho_a$, the  functional ${\cal F}$ is always {\it  local}  and is 
given by 
\begin{equation}
{\cal F}(\{\rho(x)\}|\rho_a,\rho_a)= \int_0^1 dx  \Big[ f(\rho(x)) - f(\rho_a) - (\rho-\rho_a) f'(\rho_a) \Big]
\label{F7}
\end{equation}
where $f(\rho)$ is the free energy per unit volume at density $\rho$, defined as 
$f(\rho)= - \lim_{V\to \infty}\log Z(V,V \rho)/V$ where $Z(V,N)$ is the partition function of a system  of $N$ particles at equilibrium  in a volume $V$.

In the $\rho_a \neq \rho_b$ case,  the large deviation functional ${\cal F}$ can  therefore be thought as a possible   generalization  of the concept of free energy to non-equilibrium systems.

As soon as the system is out of equilibrium ($\rho_b \neq \rho_a$) the large deviation functional  ${\cal F}$  becomes {\it non-local}.
This is already visible  in the expansion of ${\cal F}$ in 
powers of $\rho_a-\rho_b$, obtained by replacing $F(x)$ by its expansion (\ref{rhoa-rhob-small}) into 
(\ref{F2}) 
\begin{eqnarray}
\fl {\cal F}(\{\rho(x)\}|\rho_a,\rho_b)  =   \int_0^1 dx  B( \rho(x),\rho^*(x)) 
\label{Fab} \\
\fl \ \  
\ \ \ \ \  \ \  +  {(\rho_a - \rho_b)^2 \over [\rho_a (1- \rho_a)]^2}
\left[\int_0^1 dx \int_x^1 dy\: x(1-y)
  \big(\rho(x)-\rho^*(x)\big)
  \big(\rho(y)-\rho^*(y)\big) \right]
 + O(\rho_a-\rho_b)^3  
\nonumber 
\end{eqnarray}
where
the average  profile $\rho^*(x)$ is given by 
\begin{equation}
 \rho^*(x) = (1-x)  \rho_a + x \rho_b  . 
\label{rhostar}
\end{equation}

The knowledge of the functional  ${\cal F}$ allows one to determine all the  the correlation functions:
 if one  defines the generating function ${\cal G}$ of the density
by
\begin{equation}
\exp \left[L {\cal G}(\{\alpha(x)\}|\rho_a,\rho_b) \right] = \left\langle 
\exp \left[L \int_0^1 \alpha(x) \rho(x) dx  \right]  \right\rangle 
\label{Gdef}
\end{equation}
where $\alpha(x)$ is an arbitrary function and $\langle . \rangle$ denotes  an average over the  profile  $\rho(x)$ in the
steady state, 
  it is clear from (\ref{continu}) that, for large $L$,
$\cal G$ is the Legendre transform of $\cal F$
\begin{equation}
{\cal G}(\{\alpha(x)\}|\rho_a,\rho_b)  = \max_{\{\rho(x)\}}  \left[ \int_0^1 \alpha(x)
\rho(x) dx   -{\cal F} (\{\rho(x)\}|\rho_a,\rho_b) \right]   \ . 
\label{legendre}
\end{equation}
 By taking derivatives of (\ref{Gdef}) with respect to $\alpha(x)$ one can then get 
all the correlation functions. In particular
\begin{equation}
\langle \rho(x)  \rho(y) \rangle_c \equiv
\langle \rho(x)  \rho(y) \rangle
-\langle \rho(x) \rangle \langle  \rho(y) \rangle
= {1 \over L} \left. {\delta^2 {\cal G} \over \delta \alpha(x) \;  \delta \alpha(y)}\right|_{\alpha(x)=0} 
\label{2pt}
\end{equation}
A direct consequence of (\ref{Gdef}) is that all the  $k$-point connected correlation functions  are long ranged and scale like $L^{1-k}$ (see \cite{Spohn,DLS5,derrida-phys-rep}).

\section{The  steady state }

From the  definition of the SSEP,  if
$\tau_i$ is a binary variable   with $\tau_i=1$ when   site $i$ is
occupied  and $\tau_i=0$ when it is   empty, one can  write the time evolution of the average occupation $\langle
\tau_i \rangle$
\begin{eqnarray}
{d \langle \tau_1 \rangle \over dt } = & 2  \rho_a -  3 
\langle \tau_1 \rangle + \langle \tau_2 \rangle
\nonumber \\
{d \langle \tau_i \rangle \over dt } = &
\langle \tau_{i-1} \rangle -2  \langle \tau_i \rangle +  \langle
\tau_{i+1} \rangle                  \ \ \ \  \ \ {\rm for} \ \  2 \leq i
\leq L-1                     \label{evolution}                                     \\
{d \langle \tau_L \rangle \over dt } = &
\langle \tau_{L-1} \rangle - 3   \langle \tau_L \rangle +
2 \rho_b    \  . \nonumber                 \end{eqnarray}
The steady state density profile (obtained by writing that ${d
\langle \tau_i \rangle \over dt } =0$) is \cite{DLS2}
\begin{equation}
\langle \tau_i \rangle = { (L+ 1 - 2 i) \rho_a + (2 i -1) \rho_b
 \over 2 L  } \; \label{profile}
\end{equation}
For large $L$, with $i = L x$, one recovers the average density profile  (\ref{rhostar}).

In a similar way one can  then write down the equations which govern the time evolution of the two point function or higher correlations. 


\section{The matrix ansatz for the SSEP}
For the SSEP, one can then write down the steady state equations  satisfied by higher
and higher correlation functions, but solving these equations becomes quickly
complicated.

The matrix ansatz \cite{DEHP,Der,BE} gives an algebraic way of calculating exactly the weights of
all the configurations in the steady state:
in \cite{DEHP} it was shown that, in the steady state,  the probability of  a microscopic
configuration
$\{ \tau_1, \tau_2, ...\tau_L\}$ can be written as the matrix element of
a product of $L$ matrices
\begin{equation}
{\rm Pro}(\{ \tau_1, \tau_2, ...\tau_L\}) = {\langle  \rho_a | X_1 X_2 ... X_L |\rho_b 
\rangle \over \langle \rho_a | (D+E)^L |\rho_b \rangle }
\label{matrix}
\end{equation}
where the matrix $X_i$ depends on the occupation  number  $\tau_i$ 
\begin{equation}
X_i =  \tau_i  D + (1 - \tau_i) E \ ,
\end{equation}
and the matrices $D$,  $E$  and the vectors $\langle \rho_a|, |\rho_b \rangle$ satisfy the following algebraic rules
\begin{eqnarray}
&& DE-ED= D+E \nonumber \\
&& \langle \rho_a |\;  2 \;  [ \rho_a E - (1-\rho_a)D] = \langle \rho_a| \label{algebra} \\
&&   2\;  [ (1-\rho_b) D - \rho_b E]\; | \rho_b \rangle = | \rho_b \rangle  \; . \nonumber
\end{eqnarray}

A priori one should construct the matrices $D$ and $E$ (which might be
infinite-dimensional) and the vectors
$\langle \rho_a|$ and $| \rho_b \rangle$ satisfying (\ref{algebra}) to calculate the
weights of the microscopic configurations.
However these weights do not depend on the particular representation
chosen and can be calculated directly \cite{DEHP,derrida-phys-rep} from (\ref{algebra}).

One can  calculate, using (\ref{algebra}),  the average density profile
\begin{equation}
 \langle \tau_i \rangle = { \langle \rho_a | (D+E)^{i-1} D (D+E)^{L-i} |\rho_b 
\rangle \over \langle \rho_a  | (D+E)^{L} |\rho_b  
\rangle }
\label{taui}
\end{equation}
as well as all the correlation functions and  recover 
(\ref{profile}).
One can also show that 
\begin{equation}
\frac{\langle\rho_a |(D+E)^L|\rho_b\rangle}
       {\langle\rho_a |       \rho_b \rangle}=
  \frac{L!}{(\rho_a-\rho_b)^L} \ .
\label{norm}
\end{equation}
(This formula is easy to derive by noticing that,  for a system of size $L$, the average current is given, according to  (\ref{profile}),  by $\langle \tau_i - \tau_{i+1} \rangle = (\rho_a - \rho_b)/L$ 
but is also given,  according to (\ref{taui},\ref{algebra}), by the ratio 
 $ \langle\rho_a |(D+E)^{L-1}|\rho_b\rangle/\langle\rho_a |(D+E)^L|\rho_b\rangle$). 
 \section{Additivity}
As in (\ref{matrix})  the weight of each configuration is written as
the matrix element of a product of $L$ matrices, one can try to insert at
a position $L_1$ a complete basis in order to relate the properties of a
lattice of $L$ sites to those of two subsystems of sizes $L_1$ and
$L-L_1$.

If one defines, for arbitrary $\rho$,  left and right vectors $\langle \rho|$ and $|\rho\rangle$, which satisfy
\begin{eqnarray}
 && \langle \rho| \; 2 \;  [\rho E-(1-\rho)D]=   \langle  \rho|
\nonumber \\
 && 2 \; [ (1-\rho ) D- \rho E ]  \; | \rho \rangle =  | \rho
\rangle  \label{eigen}
\end{eqnarray}
(note that in general $\langle \rho | \rho' \rangle \neq 0$),
it is  possible to show, using 
$DE-ED=D+E$  as in 
  (\ref{algebra}) and 
 the property (\ref{eigen}),  that for $\rho_b <
\rho_a$
\begin{equation}
  \frac{\langle \rho_a|Y_1 Y_2|\rho_b\rangle}
       {\langle \rho_a|       \rho_b\rangle}
  =  
 \oint\displaylimits_{\rho_b<|\rho|<\rho_a} \frac{d\rho}{2i\pi}\:
    \frac{(\rho_a-\rho_b)}{(\rho_a-\rho)(\rho-\rho_b)} \:
    \frac{\langle \rho_a|Y_1|\rho\rangle}
         {\langle \rho_a|    \rho\rangle} \:
    \frac{\langle \rho|Y_2|\rho_b\rangle}
         {\langle \rho|    \rho_b\rangle}
\label{fermeture}
\end{equation}
where $Y_1$ and $Y_2$ are arbitrary polynomials of matrices $D$ and $E$.
(To prove (\ref{fermeture}),
 one can  first prove it, using  (\ref{norm}),  
  for $Y_1$  of
the form $[\rho_a E-(1-\rho_a)D]^{m_1} [D+E]^{n_1}$ and $Y_2$ of the form
$[D+E]^{n_2} [ (1-\rho_b ) D- \rho_b E ]^{m_2}$. Then one can  show, using
$DE-ED=D+E$, that any polynomial $Y_1$ or $Y_2$ can be reduced to a finite
sum of such terms).
 \\ \ 
By choosing for $Y_1$ the sum  over the weights of  all  configurations with $ l r_1$ occupied sites in the first a box, ... $l r_k$ occupied sites in the $k$th box, and  for $Y_2$  the sum over all configurations with $ l r_{k+1} $ occupied sites in the $k+1$th   box, ... $l r_n$ occupied sites in the $n$th box, one can show, using (\ref{matrix},\ref{norm},\ref{fermeture})   that 
\begin{eqnarray}
\fl
{\rm Pro}_{L}(r_1,... r_n|\rho_a,\rho_b) =
 \oint\displaylimits_{\rho_b<|\rho|<\rho_a} \frac{d\rho}{2i\pi}\: &  { L_1! \  (L-L_1)!  \over L!}   {(\rho_a - \rho_b)^{L+1} \over  (\rho_a - \rho)^{L_1+1} \  (\rho - \rho_b)^{L-L_1+1}} \times  
\label{PPP}
 \\ \nonumber 
 & \ \ \  {\rm Pro}_{L_1}(r_1,... r_k|\rho_a,\rho) 
 \times {\rm Pro}_{L-L_1}(r_{k+1},... r_n|\rho,\rho_b) 
\end{eqnarray}
where $L_1= k l$.
This formula, which is exact for arbitrary system sizes, relates the properties of two disconnected subsystems of sizes $L_1$ and $L-L_1$ to those of a single system of size $L$.
 \\ \ 

If $L_1=L x$, one then gets (\ref{finite}) for large $L$
\begin{eqnarray}
\fl {\cal F}_n(r_1, r_2, ...r_n|\rho_a,\rho_b)
= & \max_{\rho_b < F < \rho_a} \Big[
x{\cal F}_k(r_1,  ...r_k|\rho_a,F)
+(1-x){\cal F}_{n-k}(r_{k+1},  ...r_n|F,\rho_b)
\nonumber \\
& +x \log \left({\rho_a -F \over x}\right)
+(1-x) \log \left({F-\rho_b  \over 1-x}\right)
-\log(\rho_a-\rho_b) \Big]
\end{eqnarray}
which follows from   (\ref{PPP}) by a saddle point method (as in (\ref{PPP}) the integration contour is perpendicular to the real axis, the value $F$  of $\rho$ which maximizes  the integrand along  the contour becomes a minimum as $\rho$ varies along the real axis).
If one repeats the same procedure $n$ times, one gets
\begin{equation}
\label{Fn}
\fl {\cal F}_n(r_1, r_2, ...r_n|\rho_a,\rho_b)
=
 \max_{\rho_a=F_{0} >   ..  >F_k > ..> F_n=\rho_b}
{1 \over n}\sum_{k=1}^n{\cal F}_1(r_k|F_{k-1},F_k) + \log \left( {(F_{k-1}-F_k)n \over \rho_a - \rho_b}\right)
\end{equation}
For large $n$, as $F_k$ is monotone, the difference $F_{k-1}-F_k
$ has to be  small for almost all $k$ and one can replace ${\cal F}_1(r_k|
F_{k-1},F_k) $ by its equilibrium value ${\cal F}_1(r_k|F_{k},F_k) = B(r_k,F_k)$ (see (\ref{F4})).
 If one the writes  $F_k$ as a function of $k/n$
\begin{equation}
F_k = F\left({k \over n} \right)
\end{equation}
 (\ref{Fn}) becomes  
\begin{equation}
{\cal F}(\{\rho(x)\}|\rho_a,\rho_b)= \max_{F(x)}\int_0^1 dx  \ \left[ B(\rho(x),F(x)) + \log{F'(x)
\over \rho_b - \rho_a}  \right]
\label{F5}
\end{equation}
where the maximun is over all the monotone  functions $F(x)$ which satisfy
$F(0)=\rho_a$ and $F(1)=\rho_b$. Writing the equation satified by the optimal $F(x)$ in (\ref{F5}) leads to (\ref{F3}) and this completes the derivation of   (\ref{F2},\ref{F3}).
 \\ \ \\

\section{The macroscopic fluctuation theory}
\label{mft}

For a general diffusive one dimensional system  of length $L$, in contact with a left reservoir at density $\rho_a$ and a right reservoir at density $\rho_b$,
the average current and the  fluctuations of this current  
 near equilibrium can be characterized by  
two quantities $D(\rho)$ and $\sigma(\rho)$ defined by  
\begin{equation}
\lim_{t \to \infty}{\langle Q_t \rangle \over t}=  {D(\rho) \over L} (\rho_a - \rho_b)
\ \ \ \ \ \ \ {\rm for \ } 
\ (\rho_a - \rho_b) \ \   {\rm small}
\label{Ddef}
\end{equation}
\begin{equation}
\lim_{t \to \infty}{\langle Q_t^2 \rangle \over t}  = {\sigma(\rho) \over L}
\ \ \ \ \ \ \ {\rm for \ } 
\ \rho_a = \rho_b 
\label{sigmadef}
\end{equation}
where $Q_t$ is the total number of particles transferred from the left
reservoir to the system during time $t$.
Starting from the hydrodynamic large deviation theory \cite{KOV,Spohn,KL}
Bertini, De Sole, Gabrielli, Jona-Lasinio and Landim
\cite{BDGJL1,BDGJL2,BDGJL2a} have developed a   general
approach, {\it the macroscopic fluctuation theory}, to calculate the large deviation functional $\cal F$ of the density (\ref{continu}) in the 
non-equilibrium steady state of a  diffusive system in contact with two
reservoirs as in figure \ref{ssep}.

For diffusive systems (such as the SSEP),
the density $\rho_i(t)$ near site  $i$ at time $t$ and
 the total flux $Q_i(t)$ flowing through position $i$ between time $0$ and time $t$  are, 
for a large system of size $L$
and for times of order $L^2$, scaling functions of the form
\begin{equation}
 \rho_i(t) =  \widehat \rho \left( {i\over L}, {t \over L^2} \right) \; , 
\qquad {\rm and} \qquad
 Q_i(t) = L \widehat Q \left( {i\over L}, {t \over L^2} \right) \; 
\end{equation}
(Note that, due to the conservation of the number of particles $\rho_i(t)-\rho_i(0) = Q_i(t)-Q_{i+1}(t)$, the scaling form of $\rho_i(t)$ implies the scaling form of $Q_i(t)$).
If one  introduces the instantaneous (rescaled) current defined  by
\begin{equation}
\widehat j(x,\tau) = {\partial \widehat Q (x,\tau) \over \partial \tau}
\end{equation}
 the conservation of the number of particles implies that
\begin{equation} {
\partial \widehat{\rho}(x, \tau) \over \partial \tau}=
-{\partial^2 \widehat{Q} (x,\tau) \over \partial \tau \partial x }
= -{\partial \widehat{j}(x,\tau) \over   \partial x } \, .
\label{conservation}
\end{equation}

The {\it macroscopic fluctuation theory} \cite{BDGJL1,BDGJL2,BDGJL2a} starts from
the probability of observing
a certain density profile $\widehat \rho \left( x, \tau \right)$ and 
current profile $\widehat j \left( x, \tau \right)$
over the rescaled time interval $\tau_1 < \tau < \tau_2$
\begin{equation}
\label{eq: dev exp}
\fl {\cal P}_{\tau_1,\tau_2} \Big( \{\widehat \rho(x,\tau), \widehat j(x,\tau)\}  \Big)  \sim \exp
\left[ - L \int_{\tau_1}^{\tau_2}
d \tau' \int_0^1 dx {\left[\widehat j(x,\tau') + D({\widehat \rho(x,\tau'}))
{\partial {\widehat \rho(x,\tau')} \over \partial x}\right]^2 \over 2
\sigma(\widehat \rho(x,\tau'))} \right]
\end{equation}
 where the current $\widehat j(x,s)$ is related to the density profile $\widehat \rho(x,s)$
by the conservation law (\ref{conservation}) 
and
the functions $D(\rho)$ and $\sigma(\rho)$ are defined by
(\ref{Ddef},\ref{sigmadef}).   The physical meaning of (\ref{eq: dev exp})
is that  the system is {\it locally} close to equilibrium  and that the fluctuations of the local currents  are Gaussian with
 averages and  variances given by (\ref{Ddef},\ref{sigmadef}).
 
Then   to calculate the probability of  observing a density profile $\rho(x)$ in the steady state, at time $\tau$, 
 one has to find  how this profile  is produced. For large $L$, this probability (\ref{eq: dev exp})
is dominated  by  the  optimal path $\{\widehat{\rho}(x,s),\widehat{j}(x,s)\}$ for $-\infty < s < \tau$  in
the space of density and current profiles 
 and
\begin{equation}
{\rm Pro}_L(\{\rho(x)\}|\rho_a,\rho_b) \sim \max_{ \{\widehat{\rho}(x,s), \widehat{j}(x,s) \} }
{\cal P}_{-\infty,\tau}\Big(\{ \widehat{\rho}(x,s), \widehat{j}(x,s) \}\Big)
\label{ProBertini}
\end{equation}
which goes from the average steady state  profile $\rho^*(x)$  (given by (\ref{rhostar}) for the SSEP) to the desired profile $\rho(x)$
\begin{equation}
 \widehat{\rho}(x,-\infty) = \rho^*(x)  \ \ \ \ \ ; \ \ \ \
 \widehat{\rho}(x,\tau) = \rho(x) \ . \label{bc}
\end{equation}
This means that the  functional ${\cal F}$ of the density (\ref{continu}) is given by
\begin{equation}
\fl {\cal F}(\{\rho(x)\}|\rho_a,\rho_b)= \min_{\{ \widehat{\rho}(x,s), \widehat{j}(x,s) \}}
\int_{-\infty}^{\tau}
d \tau' \int_0^1 dx { \left[\widehat{j}(x,\tau') + D({\widehat{\rho}(x,\tau'}))
{\partial {\widehat{\rho}(x,\tau')} \over \partial x}\right]^2 \over 2
\sigma(\widehat{\rho}(x,\tau'))} 
\label{FBertini}
\end{equation}
where the density  and the current profiles satisfy  the conservation law (\ref{conservation})  and the  boundary conditions (\ref{bc}).

Finding this optimal path   $\widehat{\rho}(x,s),  \widehat{j}(x,s)$ with
the boundary conditions (\ref{bc})  is usually a hard problem. 
Bertini et al  \cite{BDGJL1} were however able to write an  equation satisfied by
${\cal F}$: as (\ref{FBertini}) does not depend on $\tau$ (because the probability of producing a certain deviation $\rho(x)$ in the steady state does not depend on the time $\tau$ at which 
this deviation occurs),
 one can isolate in the integral (\ref{FBertini}) the contribution of the last time interval $(\tau- \delta \tau, \tau)$ and (\ref{FBertini})   becomes
\begin{equation}
\fl {\cal F}(\{\rho(x)\})= \min_{ \delta \rho(x), j(x)} \left[ 
{\cal F}(\{\rho(x)-\delta \rho(x)\})  
+ \delta  \tau \int_0^1 dx { \left[j(x) + D(\rho(x))
\rho'(x) \right]^2 \over 2
\sigma(\rho(x))}
\right]
\label{FBertini-bis}
\end{equation}
where $\rho(x)-\delta \rho(x)=\widehat{\rho}(x,\tau-d\tau)$
and $j(x) = \widehat{j}(x,\tau)$. 
Then if one defines $U(x)$ by
\begin{equation}
U(x) = {\delta {\cal F}(\{\rho(x)\}) \over \delta \rho(x)}
\label{Udef}
\end{equation}
and one uses the conservation law
$\delta \rho(x)= -{dj(x) \over dx} d \tau$,
one should have according to (\ref{FBertini-bis})
that the optimal current $j(x)$ is given by
\begin{equation}
j(x)= -D(\rho(x)) \rho'(x) + \sigma(\rho(x)) U'(x) \ .
\label{j-bertini}
\end{equation}
 Therefore "starting" with $\widehat{\rho}(x,\tau)=\rho(x)$  at time $\tau$ and using the
time evolution  (for $-\infty < s < \tau$)
\begin{equation}
 {d \widehat{\rho}(x,s) \over ds} = - {d \widehat{j}(x,s) \over dx} 
\label{dyn}
\end{equation}
with $\widehat{j}$ related to $\widehat{\rho}$ as in  (\ref{j-bertini}) one
should get  the whole time dependent optimal profile
$\widehat{\rho}(x,s)$ which converges to $\rho^*(x)$ in the limit $s \to
-\infty$. The problem of course is that  ${ \cal F}$ is in general not known and so is $U(x)$  defined in (\ref{Udef}).

One can write from 
 (\ref{FBertini-bis})   (after an integration by parts and using the fact that $U(0)=U(1)=0$ if $\rho(0)=\rho_a$ and $\rho(1)=\rho_b$) the equation satisfied by $U'(x)$ 
\begin{equation}
 \int_0^1 dx \left[ \left({D \rho' \over \sigma} - U' \right)^2 - \left({D \rho' \over \sigma  }\right)^2  \right] {\sigma \over 2}=0
\label{hamilton-jacobi}
\end{equation}
which is the Hamilton-Jacobi equation of Bertini et al  \cite{BDGJL1}. 
For general $D(\rho)$ and $\sigma(\rho)$ one does not know how to find
the solution $U'(x)$ of (\ref{hamilton-jacobi})
 for an arbitrary $\rho(x)$ and thus
 one does not know how to get   a more explicit expression of the large deviation function ${\cal F}(\{\rho(x)\})$.

One can however  check rather easily whether a given expression of ${\cal F}$
satisfies (\ref{hamilton-jacobi}) since $U'(x)$ can be calculated from 
(\ref{Udef}). For the SSEP one gets
from (\ref{F5},\ref{Udef})
\begin{equation}
U(x) = \log \left[ {\rho(x) (1- F(x)) \over (1-\rho(x))F(x)} \right]
\end{equation}
with $F(x)$ related to $\rho(x)$  by (\ref{F3}).
One can then check that (\ref{hamilton-jacobi}) is indeed satisfied using
the known expressions of $D=1$ and $\sigma= 2 \rho(1-\rho)$ for the SSEP
\cite{derrida-phys-rep} (using the fact that $F''(0)=F''(1)=0$
which is a consequence of the fact that $\rho(0)=F(0)=\rho_a$, $\rho(1)=F(1)=\rho_b$, and of (\ref{F3})).

In fact  when
${\cal F}$ is known, one can obtain the whole optimal path
$\widehat{\rho}(x,s)$ from the evolution (\ref{dyn}) with $\widehat{j}$  related to $\widehat{\rho}$ by (\ref{j-bertini}) which becomes for the SSEP
\begin{equation}
\label{j-bertini1}
\widehat{j}(x,s)= - {d \widehat{\rho}(x,s) \over dx} + 
\sigma (\widehat{\rho}(x,s))   \log \left[ {\widehat{\rho}(x,s)  (1- \widehat{F}(x,s))  \over (1-\widehat{\rho}(x,s))  \widehat{F}(x,s) } \right]
\end{equation}
where $\widehat{F}$ is related to $\widehat{\rho}$ by (\ref{F3}).
For  (\ref{F2},\ref{F3}) to coincide with (\ref{FBertini}), the
optimal profile $\widehat{\rho}$ evolving according to (\ref{dyn}) 
should  converge to $\rho^*(x)$ as $s \to -\infty$.
One can check  that this evolution (\ref{dyn}) of $\widehat{\rho}(x,s)$  for this current (\ref{j-bertini1}) is equivalent
to the following evolution \cite{BDGJL2} of $\widehat{F}$ 
\begin{equation}
\label{F-bertini}
{d \widehat{F}(x,s) \over ds} = - {d^2 \widehat{F}(x,s) \over d x^2}
\end{equation}
where $\widehat{F}$ is related to $\widehat{\rho}$ by (\ref{F3}).
Clearly (\ref{F-bertini}) is a diffusion equation. Because  $F(0)=\rho_a$, $F(1)=\rho_b$ and  because of the minus sign in (\ref{F-bertini}),   $\widehat{F}(x,s) \to \rho^*(x)$ as $s \to - \infty$. Therefore, due to (\ref{F3}),  the density $\widehat{\rho}(x,s)
\to \rho^*(x)$ as $s \to - \infty$. Thus (\ref{dyn},\ref{j-bertini1})
 do give the optimal path  in (\ref{FBertini}) with
the right boundary conditions (\ref{bc})
and   (\ref{FBertini}) coincides for the SSEP with the prediction
(\ref{F2},\ref{F3}) of the matrix approach.
From (\ref{F-bertini},\ref{F3}) one can show  that the time evolution of a deviation $\ \widehat{\rho}(x,s)$, when it is produced is given, for small $\rho_a-\rho_b$, by
 \begin{equation}
\label{rhohat}
\fl 
{d \widehat{\rho}(x,s) \over ds}  = {d^2 \widehat{\rho}(x,s) \over dx^2} -2{ (\rho_a - \rho_b) \over \rho_a(1-\rho_a)}(1-2 \widehat{\rho}(x,s) )  {d \widehat{ \rho}(x,s) \over ds}  + O\left((\rho_a-\rho_b)^2 \right) \ . 
\end{equation}
One can notice that as soon as $\rho_a \neq \rho_b$ this is not the time reversal of the way a deviation relaxes (\ref{evolution})  
 \begin{equation}
\label{rhorel}
 {d \rho(x,t) \over dt} = {d^2 \rho(x,t)\over dx^2} 
 \ .
\end{equation}
 This again is not a surprise as for non-equilibrium systems ($\rho_a \neq \rho_b$), the way a deviation is produced (\ref{rhohat})
has no reason to be the time reversal of the way it relaxes
 (\ref{rhorel}).

\section{Conclusion}
\label{conclusion}
 In addition to the two approaches discussed above to obtain (\ref{F2}-\ref{F3}), Tailleur Kurchan and Lecomte \cite{TKL} have developed a third approach based on a non-local change of variables
which   allows them to map the dynamics of  the non-equilibrium   case ($\rho_a \neq \rho_b$) onto  the dynamics of the equilibrium case ($\rho_a=\rho_b$).

Apart from the SSEP (and zero range processes for which the steady state measure is a product measure), the large deviation function
${\cal F} $ has been determined so far only for  few other cases: 
 the Kipnis Marchioro
Presutti  model \cite{KMP,BGL},  the  weakly asymmetric exlcusion process
\cite{ED,BLM}, the ABC model \cite{EKKM,CDE} on a ring for equal
densities of the three species,  driven systems \cite{Baha} in particular  the  asymmetric exclusion process \cite{DLS3,DLS4}.

An open question is whether one could use the macroscopic fluctuation  theory to find the large deviation functional ${\cal F}$
for more general diffusive systems characterized by  arbitrary functions $D(\rho)$ and $\sigma(\rho)$ defined in (\ref{Ddef},\ref{sigmadef}).

More recently, the macroscopic fluctuation theory  has   become a very powerful tool  to  calculate  the  large deviation function of the current
in  the non-equilibrium steady state  of diffusive systems \cite{BD1,BD2,BDGJL5,BDGJL6,harris,ADLV,hurtado,HG,Imparato}.
On the other hand  exact calculations of the  current fluctuations, starting from a microscopic model,  are still very difficult to do
\cite{GE,GE1,PM,simon}.
What the large deviation functional of the density  looks like, for a diffusive system,  when conditioned on the current,  remains an  open question \cite{PSS}.

Looking, by a  macroscopic or  a microscopic approach, at diffusive systems with an initial condition which is  not a steady state as in \cite{Antoine1,Antoine2},   would  be another 
interesting   direction to  pursue.

Lastly, one knows \cite{LLP,Olla2} that  mechanical systems which conserve momentum exhibit  an anomalous Fourier's law in one dimension.
What the large deviation functions  of the current or of the density become for such systems looks to me another interesting and challenging question.

 \ack
My interest for the large deviations of the density of diffusive systems started by  the works 
 done    with Joel Lebowitz and Gene Speer  in 2001-2003.
We  used, as a starting point,  the matrix  ansatz which was developed in collaboration with Martin Evans,Vincent Hakim, Vincent Pasquier,
following an earlier paper with David Mukamel and Eytan Domany at the beginning of the 1990's.
They were pursued by a series of works on  diffusive systems, in particular  with Thierry Bodineau,
Camille Enaud and Antoine Gerschenfeld.
It was for me a great pleasure and privilege to work on these problems with them. As the talk, on which this paper is based,  was delivered on the occasion of the 
 Boltzmann Medal award, I would like  to thank also  all my other collaborators or colleagues with whom I had the opportunity to  share my interest for Statistical Physics over the past 35 years.


\section*{References}
\begin{thebibliography}{00}
\bibitem{BLR}  Bonetto F,  Lebowitz J L,  Rey-Bellet L 
 {\it Fourier's law: a challenge to theorists}  2000
 Imperial College Press 128-150    (Preprint   math-ph/0002052)

\bibitem{LLP}  Lepri S,  Livi R,  Politi A  
{\it Thermal conduction in classical low-dimensional lattices} 
2003
{\it  Phys. Rep.} {\bf    377}  1-80 


\bibitem{EPR1}
 Eckmann J P,  Pillet C A,  Rey-Bellet L 
 {\it Entropy production in nonlinear, thermally driven Hamiltonian systems}
1999
{\it J.  Stat. Phys.}   {\bf 95}  305-331 

\bibitem{ST}   Sasa S I,  Tasaki H 
 {\it Steady state thermodynamics}
2006
{\it J. Stat. Phys.}    {\bf 125}  125-227 

 \bibitem{DLS1}  
 Derrida B,   Lebowitz J L,   Speer E R 
  {\it Free energy functional for nonequilibrium systems: an exactly solvable case}
2001
{\it  Phys. Rev. Lett.}   {\bf 87} 150601

\bibitem{DLS2} 
 Derrida B,   Lebowitz J L,   Speer E R 
 {\it Large deviation of the density profile in the steady state of the open symmetric simple  exclusion process}
2002
{\it  J. Stat. Phys.}   {\bf 107} 599-634 

\bibitem{derrida-phys-rep}  Derrida B 
 {\it Non-equilibrium steady states: fluctuations and large deviations of the density and of the current}
2007
{\it J. Stat.  Mech.} P07023   

 \bibitem{BDGJL1}
 Bertini L,   De Sole A,   Gabrielli D, Jona--Lasinio G,   Landim C
 {\it Fluctuations in stationary non equilibrium states of
 irreversible processes} 
2001 {\it Phys.  Rev.  Lett.} {\bf 87} 040601 

 \bibitem{BDGJL2}
  Bertini L,   De Sole A,   Gabrielli D, Jona--Lasinio G,   Landim C
 {\it Macroscopic fluctuation theory for stationary non equilibrium states}
2002 {\it J. Stat, Phys.} {\bf 107}  635-675  


 \bibitem{BDGJL2a}
  Bertini L,   De Sole A,   Gabrielli D, Jona--Lasinio G,   Landim C 
{\it Large deviation approach to non equilibrium processes in stochastic lattice gases}
2006 {\it Bull.  Braz. Math. Soc.}  {\bf  37}    611-643  


\bibitem{Richards}
 Richards P M
 {\it Theory of one-dimensional hopping conductivity and diffusion}
1977 {\it  Phys. Rev. } B {\bf 16}  1393-1409 


\bibitem{HS}  Spohn H  1991, {\it Large scale dynamics of interacting particles}
(Springer-Verlag, Berlin)

\bibitem{Liggett}  Liggett T 1999
{\it Stochastic interacting systems: contact, voter and exclusion processes},
 324 (Springer-Verlag, Berlin)


\bibitem{KL}  Kipnis C,  Landim C 1999
{\it Scaling limits of interacting particle systems} Springer 


\bibitem{SS}    Santos  J E,  Sch\"utz G M
 {\it Exact time-dependent correlation functions for the symmetric
 exclusion process with open boundary}
2001 {\it Phys. Rev. } E {\bf  64} 036107 

\bibitem{Touchette}   Touchette H  2009
 {\it  The large deviation approach to statistical mechanics}
{\it Phys. Rep.}    {\bf  478}   1-69   


 \bibitem{BDGJLb}
  Bertini L,   De Sole A,   Gabrielli D, Jona--Lasinio G,   Landim C 
{\it  Towards a nonequilibrium thermodynamics: a self-contained macroscopic description of driven
diffusive systems}
2009 {\it J. Stat. Phys. }   {\bf  135},    857-872   

\bibitem{Spohn}  Spohn H  
{\it Long range correlations for
     stochastic lattice gases in a non-equilibrium steady state}
1983 {\it J. Phys. A }{\bf 16} 4275-4291 

 \bibitem{DLS5} 
 Derrida B,   Lebowitz J L,   Speer E R
 {\it Entropy of open lattice systems}
 2007 {\it  J. Stat. Phys. }{\bf 126}
1083-1108   

\bibitem{DEHP}
 Derrida B,  Evans M R,  Hakim V, Pasquier  V
 {\it Exact solution of a 1d asymmetric exclusion model using a matrix formulation}
1993 {\it J. Phys. A }{\bf 26} 1493-1517 

\bibitem{Der} Derrida B
 {\it An exactly soluble non-equilibrium system: the asymmetric exclusion process}
1998   {\it Phys. Rep.}  {\bf 301 } 65-83 

\bibitem{BE}  Blythe R A,  Evans  M R
 {\it Nonequilibrium steady states of matrix-product form: a solver's guide}
2007 {\it J.   Phys.  A }{\bf 40}   R333-R441   

\bibitem{KOV}  Kipnis C,  Olla S,  Varadhan S R S, 
 {\it Hydrodynamics and large deviations for simple exclusion processes}
1989 {\it Commun. Pure Appl. Math.} {\bf 42} 115-137 

\bibitem{TKL}  Tailleur J, Kurchan J, Lecomte  V
 {\it  Mapping out-of-equilibrium into equilibrium in one-dimensional transport models}
2008 {\it J.   Phys.  A} {\bf 41} 505001   


\bibitem{ED}  Enaud C,  Derrida B
 {\it Large deviation functional of the weakly asymmetric exclusion process}
 2004 {\it J.  Stat. Phys.} {\bf 114}  537-562  

\bibitem{BLM}   Bertini L,  Landim C,  Mourragui M
{\it  Dynamical large deviations for the boundary driven weakly asymmetric exclusion process }
2009 {\it Ann.  Prob.} {\bf 37}   2357-2403  

\bibitem{KMP}  Kipnis C,  Marchioro C ,  Presutti E,
 {\it Heat-flow in an exactly solvable model}
1982 {\it J. Stat. Phys.} {\bf 27}  65-74 

\bibitem{BGL}     Bertini L ,    Gabrielli D,  Lebowitz J L 
{\it Large deviation for a stochastic model of heat flow}
2005  {\it J. Stat. Phys.} {\bf 121} 843-885 

\bibitem{EKKM}
 Evans M R,  Kafri Y,  Koduvely H M, Mukamel D
 {\it Phase separation in one-dimensional driven diffusive systems}
1998 {\it Phys. Rev. Lett.}  { \bf 80}  425-429  

\bibitem{CDE}  Clincy M,  Derrida B, Evans M R
 {\it Phase transitions in the ABC model}
2003 {\it Phys. Rev.}  E {\bf 67} 066115 

 \bibitem{Baha}  Bahadoran C 2010
  A quasi-potential for conservation laws with boundary conditions 
\\ Preprint math-ph/1010.3624 

\bibitem{DLS3}
  Derrida B,  Lebowitz J L,  Speer E R
{\it Exact free energy functional for a driven diffusive open stationary
 nonequilibrium system}
2002 {\it Phys. Rev. Lett. } {\bf 89} 030601  

 \bibitem{DLS4} 
  Derrida B,  Lebowitz J L,  Speer E R
 {\it Exact large deviation functional of a stationary open driven
 diffusive system: the asymmetric exclusion process}
2003 {\it J. Stat. Phys.} {\bf 110} 775-810 


\bibitem{BD1}
  Bodineau T,  Derrida B
{\it  Current fluctuations in nonequilibrium diffusive systems: an additivity
 principle }  
2004 {\it Phys. Rev. Lett.} {\bf 92}  180601 


\bibitem{BD2}
  Bodineau T,  Derrida B
{\it Distribution of current in nonequilibrium diffusive systems and phase transitions}
2005  {\it Phys. Rev.} E   {\bf 72}  066110   


 \bibitem{BDGJL5}
  Bertini L,   De Sole A,   Gabrielli D, Jona--Lasinio G,   Landim C
 {\it Current fluctuations in stochastic lattice gases}
2005 {\it Phys. Rev. Lett.} {\bf 94} 030601 


\bibitem{BDGJL6}
  Bertini L,   De Sole A,   Gabrielli D, Jona--Lasinio G,   Landim C
{\it Non equilibrium current fluctuations in stochastic lattice gases}
2006 {\it J. Stat. Phys.}  {\bf 123}   237-276  

\bibitem{harris}
Harris R J, Sch\"utz GM
{\it Fluctuation theorems for stochastic dynamics }
2007 {\it J.  Stat. Mech. }   P07020   

\bibitem{ADLV}
Appert C,  Derrida B,  Lecomte V,  Van Wijland F 
{\it  Universal cumulants of the current in diffusive systems on a ring}
2008 {\it   Phys. Rev.} E {\bf 78} 021122 
\bibitem{hurtado}
 Hurtado P I,  Garrido  P L
 {\it Test of the additivity principle for current fluctuations in a model of heat conduction}
2009 {\it Phys. Rev. Lett.} {\bf 102}  250601  


\bibitem{HG}
 Hurtado P I, Garrido P L
 {\it Current fluctuations and statistics during a
  large deviation event in an exactly solvable transport model} 
2009 {\it J. Stat.  Mech: Theory Exp.}  P02032  

\bibitem{Imparato}
 Imparato A,  Lecomte V,  van Wijland  F
 {\it Equilibriumlike fluctuations in   some boundary-driven open diffusive systems} 
2009 {\it  Phys. Rev.} E {\bf 80} 011131

\bibitem{GE}
de Gier J,  Essler F H 
{\it Bethe ansatz solution of the asymmetric exclusion process with open boundaries}
2005 {\it Phys. Rev. Lett.} {\bf 95} 240601

\bibitem{GE1}
de Gier J, Essler F H
{\it Slowest relaxation mode of the partially asymmetric exclusion process with open
boundaries} 2008 {\it J. Phys. A} {\bf 41} 485002

\bibitem{PM}
Prolhac S, Mallick K
{\it  Cumulants of the current in a weakly asymmetric exclusion process}
2009
{\it  J.   Phys.  A } {\bf 42}    175001  

\bibitem{simon}
 Simon D {\it Construction of a coordinate Bethe Ansatz for the asymmetric exclusion process with open boundaries} 2009 {\it J. Stat. Mech.}  P07017


\bibitem{PSS}
 Popkov V,  Simon D,   Sch\"utz G M  
{\it Asymmetric simple exclusion process on a ring conditioned on enhanced flux }
2010 {\it J. Stat. Mech.} P07017 

\bibitem{Antoine1}  Derrida B,  Gerschenfeld  A
{\it Current fluctuations of the one dimensional symmetric simple exclusion process with step initial Condition}
2009 {\it J. Stat. Phys.} {\bf  136} 1-15   

\bibitem{Antoine2}  Derrida B,  Gerschenfeld  A
{\it  Current fluctuations in one dimensional diffusive systems with a step initial density profile}
2009 {\it J. Stat. Phys.} {\bf  137}   978-1000   

\bibitem{Olla2} 
 Basile G,  Bernardin C,   Olla  S
{\it Momentum conserving model with anomalous thermal conductivity in
 low dimensional systems}
2006 {\it Phys.  Rev.  Lett. }{\bf 96}  204303 






\end{thebibliography}
\end{document}